\documentclass[final,3p,times]{elsarticle}
 \biboptions{comma,sort&compress}
\usepackage{here}
 \usepackage{graphicx}
  \usepackage{epsfig}
\usepackage{subfig}
\usepackage{amsmath,bm}

\makeatletter
\def\ps@pprintTitle{%
  \let\@oddhead\@empty
  \let\@evenhead\@empty
  \let\@oddfoot\@empty
  \let\@evenfoot\@oddfoot
}
\makeatother

\def\beq{\begin{equation}}
\def\eeq{\end{equation}}
\def\bea{\begin{eqnarray}}
\def\eea{\end{eqnarray}}

\def\lesssim{\ \hbox{\raise 2pt \hbox{$<$} \kern -13pt
                     \lower 3pt \hbox{$\sim$}}\ }
\def\greatersim{\ \hbox{\raise 2pt \hbox{$>$} \kern -13pt
                     \lower 3pt \hbox{$\sim$}}\ }
\def\frac#1#2{ {{#1} \over {#2} }}

\input epsf.tex
\def\desepsf(#1 width #2){\epsfxsize=#2 \epsfbox{#1}}
\def\kt{k_{\perp}}
\def\mut{\mu_\perp}
\def\mutpr{\mu_\perp^{\prime}}
\def\mutp2{\mu_\perp^{\prime 2}}

\newcommand{\nn}{\nonumber}

\begin{document}

\begin{frontmatter}

\hspace*{13.00 cm} {\small CERN-TH-2022-087} \\
\hspace*{13.55 cm} {\small IFJPAN-IV-2022-8}
\vspace*{1.4 cm} 
\title{A parton branching with transverse momentum dependent splitting functions}
\author[label1,label2,label3]{F.~Hautmann}
 \address[label1]{Universiteit Antwerpen, Elementaire Deeltjes Fysica, B 2020 Antwerpen}
  \address[label2]{CERN, Theory Department, CH 1211 Geneva}
\address[label3]{University of Oxford, Theoretical Physics Department, Oxford OX1 3PU}

\author[label4]{M.~Hentschinski}
\address[label4]{
%Departamento de Actuaria, F\'isica y Matem\'aticas, 
Universidad de las Americas Puebla, San Andr\'es Cholula, 
72820 Puebla, Mexico}

\author[label1]{L.~Keersmaekers}

\author[label5]{A.~Kusina}
\address[label5]{Institute of Nuclear Physics, Polish Academy of Sciences, 
 ul.~Radzikowskiego 152, 31-342, Krak\'ow}

\author[label5]{K.~Kutak}

\author[label1]{A.~Lelek}

\begin{abstract}
Off-shell, transverse-momentum dependent splitting functions can be defined 
from the high-energy limit of partonic decay amplitudes. 
Based on these splitting functions, we construct Sudakov form 
factors  and formulate a new parton branching algorithm. We present a first 
Monte Carlo implementation of the algorithm.  We use the numerical results 
to verify explicitly momentum sum rules for TMD parton distributions.   
\end{abstract}

\end{frontmatter}

Experimental analyses in  
high-energy physics  depend on event 
simulations performed through Monte Carlo (MC)  
generators~\cite{Buckley:2019kjt} based on parton branching methods. 
The development of 
 MC event generators is crucial for the planning of 
future experimental  programs such as 
 the High-Luminosity Large Hadron Collider 
(HL-LHC)~\cite{Azzi:2019yne}, the proposed forward physics 
facility~\cite{Feng:2022inv}  and hadron-electron 
facility~\cite{LHeC:2020van} at the HL-LHC, the 
Electron Ion Collider (EIC)~\cite{Proceedings:2020eah} and the 
Future Circular Collider (FCC)~\cite{Mangano:2016jyj}.   

While most MC tools rely on 
the description of hadron structure provided by collinear 
parton distribution functions (PDFs)~\cite{Kovarik:2019xvh}, 
  ongoing  advances in MC generators raise   the 
 question of assessing the impact 
of a more complete description of hadron structure including 
transverse momentum dependent (TMD) 
parton distributions~\cite{Angeles-Martinez:2015sea} on MC simulations.   
In fact, several developments of the last few 
years in parton branching methods have 
 involved aspects of TMD physics. 
 This includes, for 
instance, TMD perturbative resummation 
and its matching with finite-order next-to-next-to-leading  (NNLO) 
corrections~\cite{Chen:2022cgv,Bizon:2018foh}; 
parton branching formulation of the evolution 
of TMD distributions~\cite{Hautmann:2017xtx,Hautmann:2017fcj};   
implementation of soft and collinear 
corrections in parton showers with subleading-logarithmic 
accuracy~\cite{vanBeekveld:2022zhl,Gellersen:2021eci}; 
multi-jet merging  with TMD parton showers~\cite{Martinez:2021chk,Martinez:2021dwx}.   

In this work we begin an investigation of the  
transverse momentum dependence 
 at the level of  the partonic  splitting 
 functions~\cite{Gieseke:2003rz,Collins:2003fm,Hautmann:2007uw}, an aspect  
which   has not been explored so far in  parton branching MC. 
 To this end we propose using  off-shell  TMD   
splitting functions defined from the high-energy limit of 
QCD multi-parton amplitudes 
according to the high-energy factorization method~\cite{Catani:1994sq}.  
We  construct a parton branching formalism based on the 
TMD splitting functions thus defined.

To do this, we employ the 
approach~\cite{Hautmann:2017xtx,Hautmann:2017fcj} and extend it 
to introduce new real-emission TMD splitting kernels 
and new Sudakov form factors.  This approach makes use of 
the concept of ``unitarity", commonly applied in parton-showering 
algorithms, to relate real and virtual emissions, and to express Sudakov form 
factors in terms of real-emission kernels and soft-gluon resolution scales. In this respect it  
differs from the treatment of Sudakov factors 
in 
terms of integrals over virtual emissions, used in 
several computations  based on TMD dynamics, 
e.g.~\cite{Camarda:2019zyx,Camarda:2021ict} 
and~\cite{Coradeschi:2017zzw,Accomando:2019ahs}.  
The approach we use 
is well-suited for the implementation of the 
transverse momentum dependence in real-emission kernels, 
exploiting the positivity of the 
splitting function defined through the 
method~\cite{Catani:1994sq}.  
Since the  splitting function~\cite{Catani:1994sq}, once combined with 
 factorization formulas in transverse momentum, accomplishes the 
  small-$x$ resummation in the evolution kernels~\cite{Catani:1993rn},  this study constitutes 
 a first step toward  a full generator  
 extending~\cite{Hautmann:2017xtx,Hautmann:2017fcj}   
 to the small-$x$ phase space~\cite{Monfared:2019uaj}.  
  In this Letter 
we present the branching evolution equations 
which result from this approach, and illustrate 
two numerical applications to the momentum sum rule and 
to the evolution of TMD parton distributions. An earlier 
discussion of  results from this investigation 
may be found in~\cite{Keersmaekers:2021arn}.

We will proceed as follows. We will first briefly discuss the TMD splitting 
functions. Next we will describe their parton branching implementation. We 
will finally present  MC results from the numerical solution of the 
branching equations.  
 
\begin{figure}[htb]
\begin{center} 
\includegraphics[width=5cm]{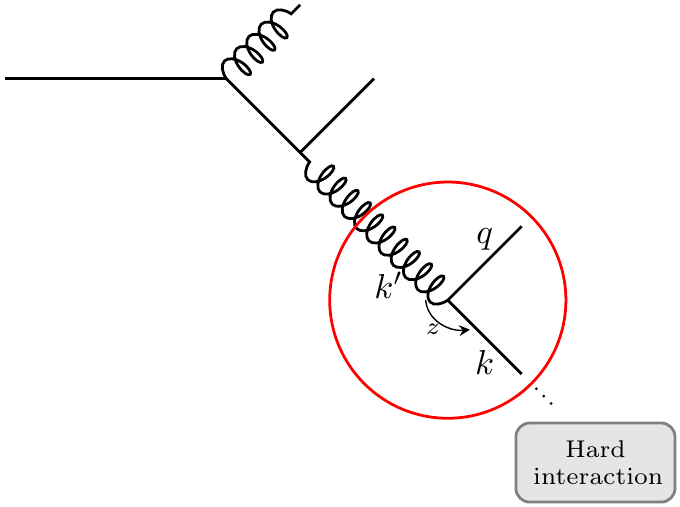}
  \caption{\it Spacelike parton cascade. %The cross on the right  
%  represents the  coupling to the hard process.
}
\label{fig1:kinem}
\end{center}
\end{figure}

Consider the initial-state (spacelike) parton cascade in Fig.~\ref{fig1:kinem}. 
We use a Sudakov parameterization of the four-momenta along the decay chain 
in terms of lightcone and transverse momenta. For 
 the gluon to quark splitting process depicted in Fig.~\ref{fig1:kinem} we    
parameterize  the  four-momenta $k$ and $k^\prime$ as  
\begin{equation} 
\label{sudkin} 
k = x p + k_T + { {k^2 + {k}_\perp^2} \over  { 2 x p \cdot {\overline p} } } \  {\overline p} \;\; \; , \;\;  \;\;\; 
k^\prime = x^\prime p + k_T^\prime + { {k^{\prime 2} + {k}_\perp^{\prime 2}} \over  
{ 2 x^\prime p \cdot {\overline p} } } \ {\overline p} \; . 
\end{equation} 
Here we use the notation 
 $v^\mu = ( v^+, v^-, {v}_\perp)$ for any four-vector, with $v^\pm = (v^0 \pm v^3) / \sqrt{2}$  lightcone 
 components and 
 ${ v}_\perp $ two-dimensional euclidean vector;   
 the reference (lightcone)    
momenta $p$ and ${\overline p}$ are 
$ p =  (\sqrt{s/2} , 0 ,  0_\perp) $, $  {\overline p} =  (0 , \sqrt{s/2} ,   0_\perp) $;   
the transverse momenta fulfill 
$ k_T^2 = - k_\perp^2$, $ k_T^{\prime 2} = - k_\perp^{\prime 2}$. 
 We define the lightcone momentum transfer at the splitting as  $z = x / x^\prime $. 
  By four-momentum conservation,    we have 
 \begin{equation} 
\label{4momcons} 
k^{\prime 2} = {q^2 \over {1-z} }  +   { k^{2}  \over z}  +  {{\tilde q}_\perp^2 \over {z(1-z)} }     \; ,   
\end{equation} 
 where $ {\tilde q}_\perp = k_\perp    - z k_\perp^\prime$.

 The splitting probability 
 for the off-shell gluon to quark splitting process in Fig.~\ref{fig1:kinem} can be 
 defined and computed  
by high-energy factorization~\cite{Catani:1994sq}  as a 
 function of the strong coupling $\alpha_s$,  lightcone momentum transfer $z$, and 
  transverse momenta $k_\perp^\prime$ and $ {\tilde q}_\perp$.
  Its explicit expression  
 is given 
  by 
\begin{equation}
  \label{eq:ktsplitt_def}
  {P}_{q g}  \left(\alpha_s, z, {k_\perp^\prime} ,  {\tilde q}_\perp \right) = 
{{\alpha_s T_F} \over { 2 \pi} }  
      { {{\tilde q}_\perp^2 z (1-z) } \over  { ({\tilde q}_\perp^2 + z(1-z){k}_\perp^{\prime 2} )^2}    }       
\left[
 \frac{{\tilde q}_\perp^2}{z(1-z)} 
+
 4 (1-2 z)  {\tilde q}_\perp\cdot { k}_\perp^\prime
 -
 4\frac{({\tilde q}_\perp\cdot { k_\perp^\prime})^2  }{{ k}_\perp^{\prime 2}} + 4 z(1-z) { k}_\perp^{\prime 2 } 
\right] ,  
\end{equation}  
where $T_F$ is the color trace invariant, $T_F = 1/ 2$.  For  
$k_\perp^{\prime 2} \ll k_\perp^{ 2}$, 
after angular average 
the TMD splitting probability in Eq.~(\ref{eq:ktsplitt_def}) 
 returns, for all lightcone momentum 
fractions $z$, the leading-order collinear splitting 
function~\cite{dglapref1,dglapref2,dglapref3}.  
 For finite $k_\perp^{\prime 2} $ such that 
$k_\perp^{\prime 2} \sim {\cal O} (k_\perp^{ 2})$,  on the other hand, Eq.~(\ref{eq:ktsplitt_def}) gives a series expansion 
 in powers $ ({k_\perp^{\prime 2}} /  {{\tilde q}_\perp^2})^n$,  with $z$-dependent coefficients. 
These finite-$k_\perp^{\prime 2} $ contributions, through 
convolution with  transverse momentum dependent 
gluon Green's functions~\cite{Catani:1994sq,Kuraev:1977fs,Balitsky:1978ic}, provide the resummation  
of the higher-order corrections to the gluon to quark splitting function 
that are logarithmically enhanced for $z \to 0$~\cite{Catani:1993rn}, 
at all orders in $\alpha_s$.   
Off-shell TMD splitting functions obtained by high-energy factorization have been 
further studied in  the  context  of 
forward Drell-Yan production  in~\cite{Hautmann:2012sh}, and have been computed for 
 all partonic channels in~\cite{Gituliar:2015agu,Hentschinski:2016wya,Hentschinski:2017ayz}.  
 These splitting functions are positive definite and interpolate consistently between the collinear limit and 
 the high-energy limit~\cite{Catani:1994sq,Gituliar:2015agu,Hentschinski:2017ayz}. 

To construct a parton branching based on TMD splitting functions, 
we extend  the method~\cite{Hautmann:2017fcj}.  We introduce the soft-gluon 
resolution scale $z_M$ to separate resolvable and non-resolvable branchings, 
and consider the branching 
evolution of the momentum weighted TMD parton distributions $  \tilde{\mathcal{A}}_a(x,\kt^2,\mu^2)$, 
where $a$ is the flavor index, $x$ is the longitudinal momentum fraction, $\kt$ is the 
transverse momentum, and $\mu$ is the evolution variable.  
We require that the resolvable branchings are described by emission kernels 
given by the TMD splitting functions.   Based on the behavior   of TMD distributions with respect to 
the resolution scale $z_M$  analyzed in~\cite{Hautmann:2017xtx,Hautmann:2019biw}, we  
also require  the evolution to be angular-ordered. For the evolution of 
 $  \tilde{\mathcal{A}}_a$ 
from scale $\mu_0$ to scale $\mu$ we write 
\begin{align}
\label{eq:ap:evol}
 \tilde{\mathcal{A}}_a(x,\kt^2,\mu^2) & =\tilde{\mathcal{A}}_a(x,\kt^2,\mu_0^2) -\int\frac{d^2\mutpr}{\pi\mutp2}F_a(\mutp2,\kt^2)\tilde{\mathcal{A}}_a(x,\kt^2,\mutp2)\Theta(\mutp2-\mu_0^2)\Theta(\mu^2-\mutp2)+ \nn
\\ &\hspace{-2.2cm} +\sum_b\int\frac{d^2\mutpr}{\pi\mutp2} \int\limits_x^{z_M}dz\tilde P^R_{ab}(z,\kt + (1-z)\mutpr,\mutpr)\tilde{\mathcal{A}}_b\left(\frac{x}{z},(\kt+(1-z)\mutpr)^2,\mu'^2\right)\Theta(\mutp2-\mu_0^2)\Theta(\mu^2-\mutp2),   
\end{align}
where  the virtual corrections 
and non-resolvable branchings are collectively 
represented by the contribution  in $F_a$ in the first line, in which $F_a$ is a kernel to be determined,  and 
the resolvable branchings are described by the term in the second line through  the 
 TMD, fully angle-dependent  splitting functions $\tilde P^R_{ab}$. 
 The explicit expressions for the functions $\tilde P^R_{ab}$ 
 are given in~\cite{Catani:1994sq,Hautmann:2012sh,Gituliar:2015agu,Hentschinski:2016wya,Hentschinski:2017ayz}.

To determine the specific form of $F_a$, we apply the 
``unitarity" approach, analogously to~\cite{Hautmann:2017fcj}. Using  
four-momentum conservation, we require that the sum over flavors of the 
normalization integrals for TMD distributions $  \tilde{\mathcal{A}}_a$ 
 is  not changed by  evolution, so that 
\begin{equation}
0=\sum_a\int_0^1dx\int  d \kt^2\tilde{\mathcal{A}}_a(x,\kt^2,\mu^2)-\sum_a\int_0^1dx\int  d \kt^2\tilde{\mathcal{A}}_a(x,\kt^2,\mu_0^2).
\label{eq:ap:conservation}
\end{equation}
Inserting  now the TMD PDF at the scale $\mu$ in the above relation, using  Eq.~\eqref{eq:ap:evol} 
and subsequently substituting  $\kt'=\kt+(1-z)\mut$ as well as  $t=x/z$,
the momentum sum rule  yields the following relation between the real splitting functions and the non-resolvable branchings:
\begin{align}
0 = \sum_b\int\frac{d^2\mutpr}{\pi\mutp2}\Theta(\mutp2-\mu_0^2)\Theta(\mu^2-\mutp2)\int\limits_0^1dt\int
    d k_{\bot}^{\prime 2}
  \left(F_b(\mutp2,\kt^{\prime 2}) - \sum_a\int\limits_0^{z_M}dz\ z\tilde P^R_{ab}(z,\kt',\mutpr)\right) \tilde{\mathcal{A}}_b(t,\kt^{\prime 2},\mutp2).
\end{align}

Therefore the sum rule allows us to fix the still missing term corresponding to non-resolvable branchings.  Introducing the angular averaged TMD
splitting functions $\bar P^R_{ba}(z,\kt^2,\mu'^2)$,
we have
\begin{align}
F_a(\mu'^2,\kt^2) =\sum_b\int\limits_0^{z_M}dz\ z\bar P^R_{ba}(z,\kt^2,\mu'^2).
\end{align}
With that we write Eq.~\eqref{eq:ap:evol} in  differential form,
\begin{align}
\frac{d\tilde{\mathcal{A}}_a(x,\kt^2,\mu^2)}{d\ln\mu^2} & = -\tilde{\mathcal{A}}_a(x,\kt^2,\mu^2)\sum_b\int_0^{z_M}dz\ z\bar P^R_{ba}(z,\kt^2,\mu^2) \notag \\
&\hspace{-1.5cm}+ \sum_b\int\frac{d^2\mutpr}{\pi}\delta(\mu^2-\mutp2) \int\limits_x^{z_M}dz\tilde P^R_{ab}(z,\kt+(1-z)\mutpr,\mutpr)\tilde{\mathcal{A}}_b\left(\frac{x}{z},(\kt+(1-z)\mutpr)^2,\mu^2\right),
\label{eq:ap:evoldiff}
\end{align}
and introduce the TMD  Sudakov form factor,
\begin{equation}
\Delta_a(\mu^2,\mu_0^2,\kt^2)\equiv\Delta_a(\mu^2,\kt^2)=\exp\left(-\sum_b\int_{\mu_0^2}^{\mu^2}\frac{d\mu'^2}{\mu'^2}\int_0^{z_M}dz\ z\bar P^R_{ba}(z,\kt^2,\mu'^2)\right). \label{eq:ap:sud}
\end{equation}
Using
\begin{equation}
\frac{d\Delta_a(\mu^2,\kt^2)}{d\ln\mu^2}=-\Delta_a(\mu^2,\kt^2)\sum_b\int_0^{z_M}dz\ z\bar P^R_{ba}(z,\kt^2,\mu^2),
\end{equation}
we arrive at
\begin{align}
\frac{d\tilde{\mathcal{A}}_a(x,\kt^2,\mu^2)}{d\ln\mu^2}
&=
\frac{1}{\Delta_a(\mu^2,\kt^2)}\frac{d\Delta_a(\mu^2,\kt^2)}{d\ln\mu^2}\tilde{\mathcal{A}}_a(x,\kt^2,\mu^2) \notag \\
& \hspace{-1cm}
+\sum_b\int\frac{d^2\mutpr}{\pi}\delta(\mu^2-\mutp2) \int\limits_x^{z_M}dz\tilde P^R_{ab}(z,\kt+(1-z)\mutpr,\mutpr)\tilde{\mathcal{A}}_b\left(\frac{x}{z},(\kt+(1-z)\mutpr)^2,\mu^2\right).
\label{eq.11}
\end{align}
By dividing Eq.~(\ref{eq.11})  by $\Delta_a(\mu^2,\kt^2)$ and  integrating over $d\ln\mu^2$, we obtain 
\begin{align}
 \tilde{\mathcal{A}}_a\left( x,\kt^2, \mu^2\right) &= 
 \Delta_a\left(\mu^2,\kt^2\right)\tilde{\mathcal{A}}_a\left( x,\kt^2, \mu_0^2\right) + \nonumber \\
& \hspace{-1.9cm} 
 \sum_b\int\frac{d^2\mutpr}{\pi\mutp2}\Theta(\mutp2-\mu_0^2)\Theta(\mu^2-\mutp2)
  \int\limits_x^{z_M }\textrm{d}z\,  \frac{ \Delta_a\left(\mu^2, \kt^2  \right)  }  
  { \Delta_a\left(\mutp2, \kt^2 \right)}  \tilde{P}_{ab}^{R}\left(z, \kt +(1-z)\mutpr, \mutpr\right) 
  \tilde{\mathcal{A}}_b\left( \frac{x}{z},  (\kt+(1-z)\mutpr)^2, \mutp2\right).
\label{eq:tmdPevolktdepSud}
\end{align}
The evolution equation (\ref{eq:tmdPevolktdepSud}) implies the 
introduction of the new 
Sudakov form factor defined in Eq.~(\ref{eq:ap:sud}) in terms of angular-averaged 
TMD splitting functions. This is one of the main results of this paper. 
It can be compared with  recent approaches~\cite{Mueller:2013wwa,Marzani:2015oyb,Nefedov:2021vvy,Hentschinski:2021lsh}  
 aiming at a combined treatment of Sudakov and small-$x$ 
contributions to parton evolution.  
The distinctive feature of the approach presented in this paper is that the treatment is done 
at the level of unintegrated, $k_\perp$-dependent splitting functions which factorize 
in the high-energy limit and control the summation of small-$x$ logarithmic contributions to 
the evolution. These  splitting functions are then used in the branching algorithm, 
where they are integrated to construct the new Sudakov factors.  In what follows we present 
a numerical MC implementation of the Sudakov factor and evolution equation, and use the 
MC results to illustrate properties of the new algorithm.

To solve the new branching equation by the MC method, 
we note that all TMD splitting functions are positive definite, 
so that the procedure~\cite{Hautmann:2017fcj} 
applies.  
Compared to~\cite{Hautmann:2017fcj}  
we however adapt the  implementation of the Sudakov form factor,  
to take account  of the  $k_\bot$-dependence,  
by making use of the veto  algorithm~\cite{Sjostrand:2006za}. 
 In the  MC code, the scale of the next
branching $\mu_i$ is generated according to the Sudakov form factor, 
\begin{equation} 
\label{eq:Rgenerat}
R=1-\frac{\Delta_a(\mu_i^2)}{\Delta_a(\mu_{i-1}^2)}\Leftrightarrow 
\mu_i^2=\Delta_a^{-1}((1-R)\Delta_a(\mu_{i-1}^2)), 
\end{equation} 
where R represents a uniformly distributed random number between zero and one.
To find the inverse of the Sudakov form factor and generate $\mu_i$, 
in the case of  collinear splitting functions  a table of
Sudakov  factors is computed and interpolation methods are used to
calculate $\mu_i$. In the case of the $k_\bot$-dependent Sudakov form factor, 
instead of extending  the 
Sudakov table by an additional dimension  
 we  use the  veto algorithm. By writing 
$\Delta_a(\mu^2,k_{\bot}^2)=\exp\left(-\int_{\mu_0^2}^{\mu^2}  ({d\mu'^2}/{\mu'^2}) g_a(\mu'^2,k_{\bot}^2)\right)$, 
with $g_a(\mu^2,k_{\bot}^2)=\sum_b\int_0^{z_M}dz\ z\ \bar P^{R}_{ba}(z,k_\bot^2,\mu'^2)$ 
the differential branching probability, in the veto algorithm one proposes to find a function $g_a^{\prime}$ such that
$g_a^{\prime}(\mu^2)\geq g_a(\mu^2,k_{\bot}^2)$ for all $\mu$ and $k_{\bot}$ and proceeds in the following way:
\begin{enumerate}
\item start with $j=0$, $p_{j=0}^2=\mu_{i-1}^2$,
\item add one to j. Select $p^2_{j}>p^2_{j-1}$ according to 
$R=1-\exp\left(-\int_{p_{j-1}^2}^{p_j^2}  ({dp'^2}/{p'^2}) g_a^{\prime}(p'^2)\right)$,
\item if $g(p^2_j)/g^{\prime}(p^2_j)\leq (\rm{newly\;generated})\; R$ go to 2,
\item else: $\mu_i^2=p^2_j$ is the generated scale.
\end{enumerate}
The function $g^{\prime}$ is usually chosen to have a known inverse function,
but  for our application we choose 
$g_a^{\prime}(\mu^2)=\sum_b\int_0^{z_M}dz\ z\ (P^{R}_{ba}(z)+h_{ba}(z))$, where the first 
term is the integral of the collinear splitting function and  the function
$h_{ba}$ is  added to ensure that  $g^{\prime}$ is larger
than $g$ for all values of $k_\bot$ and $\mu$. This function $g^{\prime}$ does
not have a known inverse, but has one  variable less than the function
$g$ (no $k_\bot$) and can be dealt with by using tables, similar to the
ones from the collinear Sudakov form factor. It is also close to $g$  
which makes the algorithm efficient.
The other variables of the splittings can be 
simply computed according to the  method~\cite{Hautmann:2017fcj}  
but with TMD splitting functions.

\begin{figure}[h!]

\begin{center} 

    \subfloat{
    \includegraphics[width=0.4\textwidth]{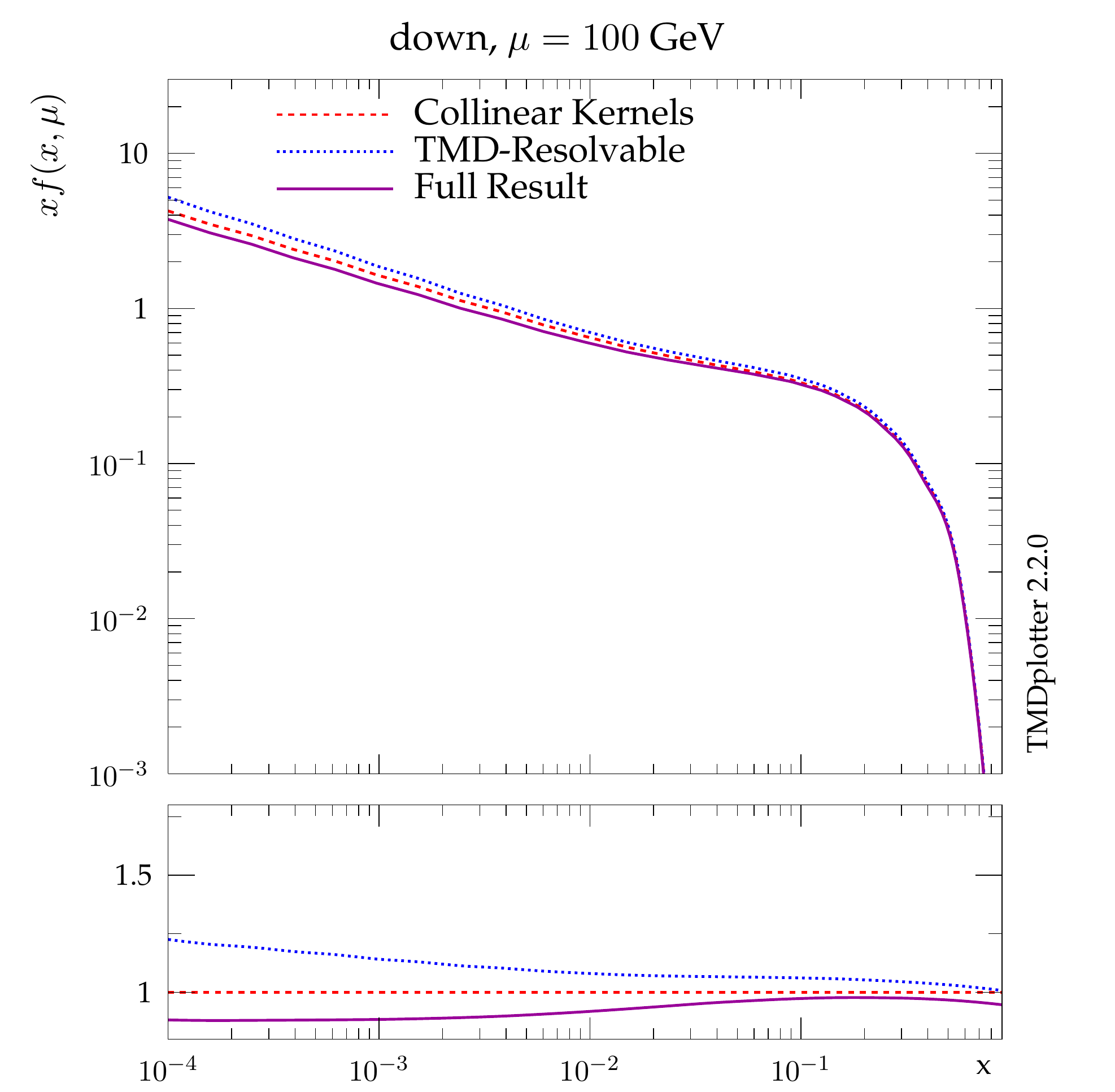}
    }
    \qquad
    \subfloat{
    \includegraphics[width=0.4\textwidth]{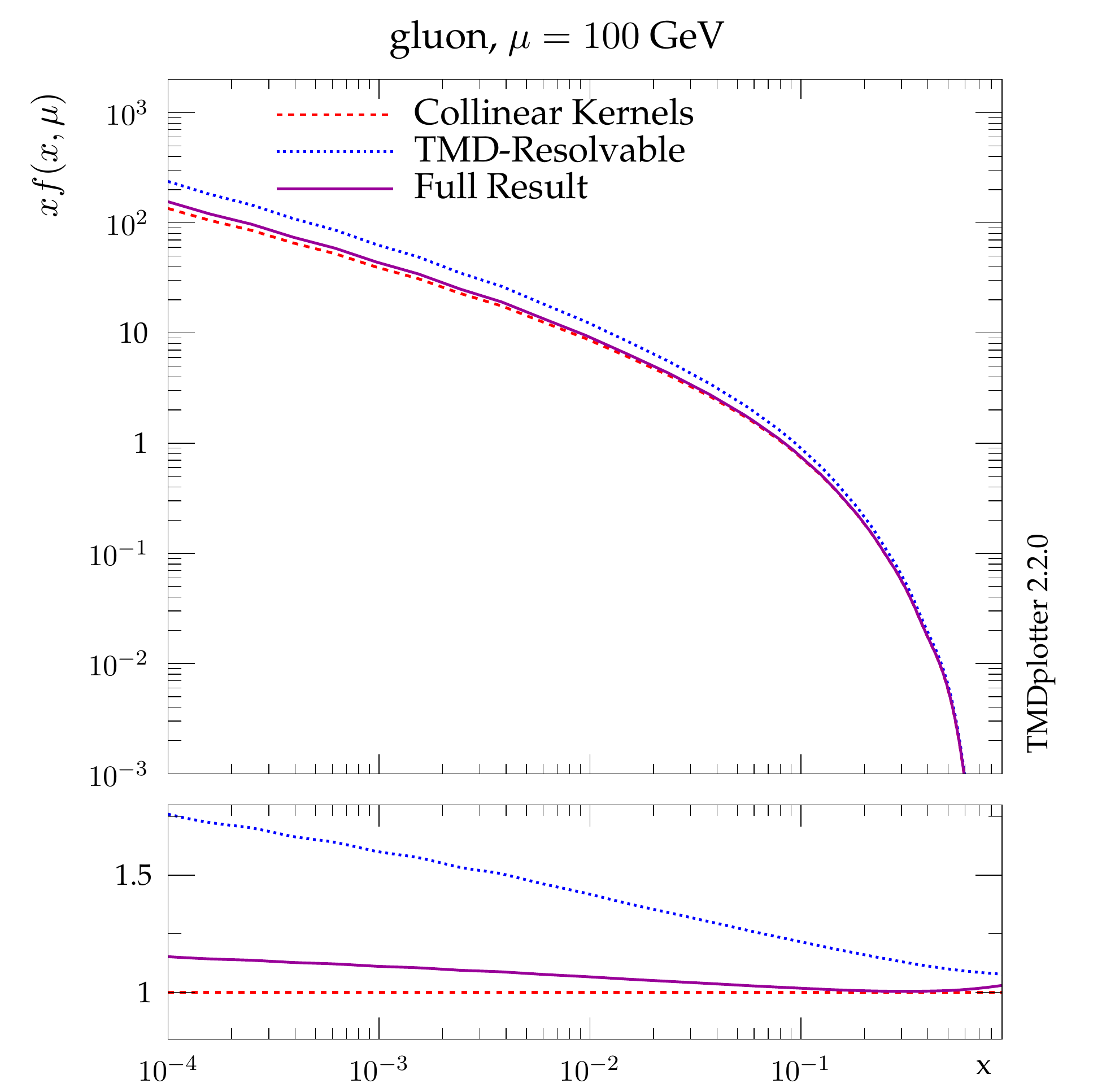}
    }
    \qquad
        \subfloat{
    \includegraphics[width=0.4\textwidth]{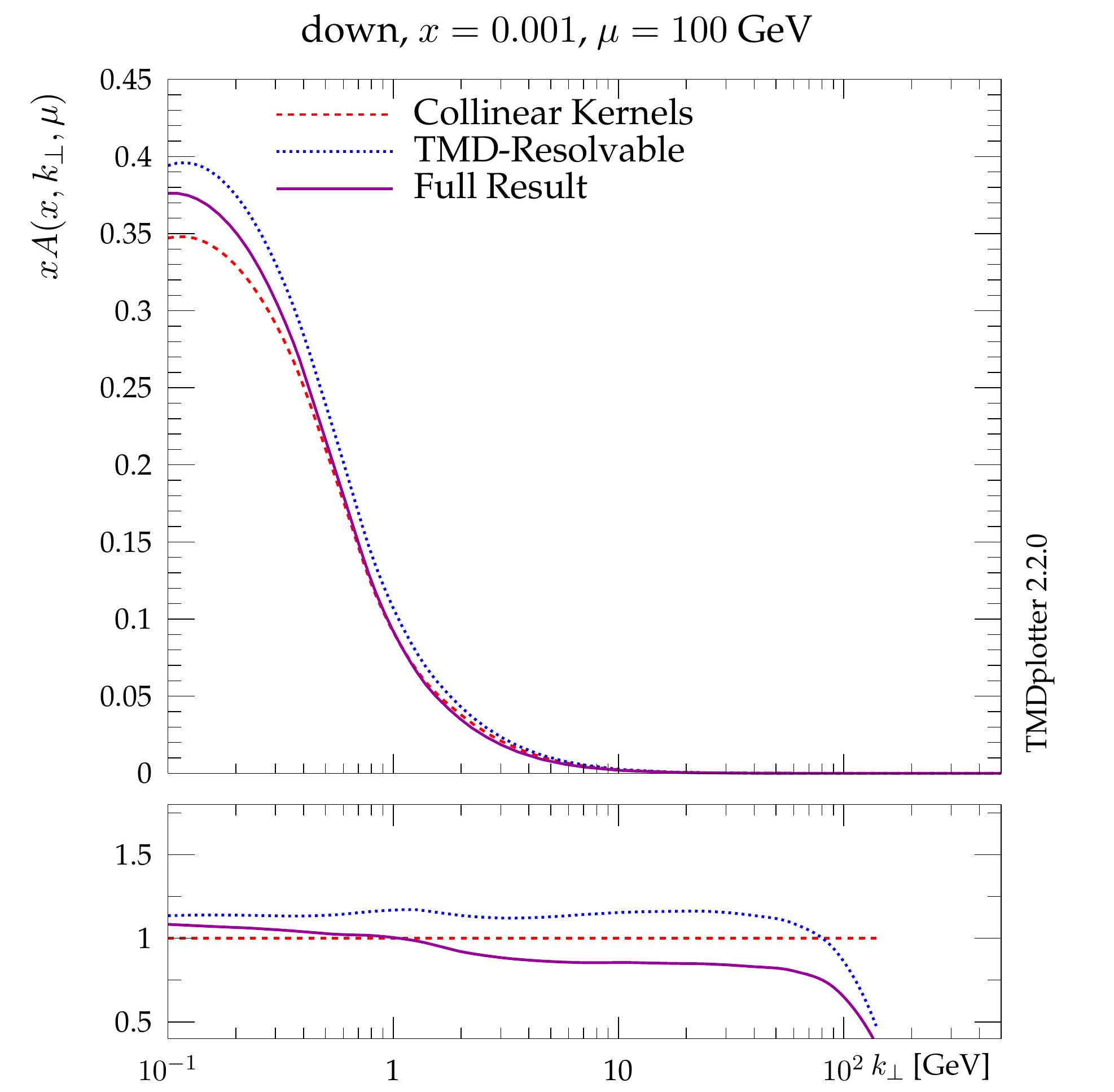}
    }
    \qquad
        \subfloat{
    \includegraphics[width=0.4\textwidth]{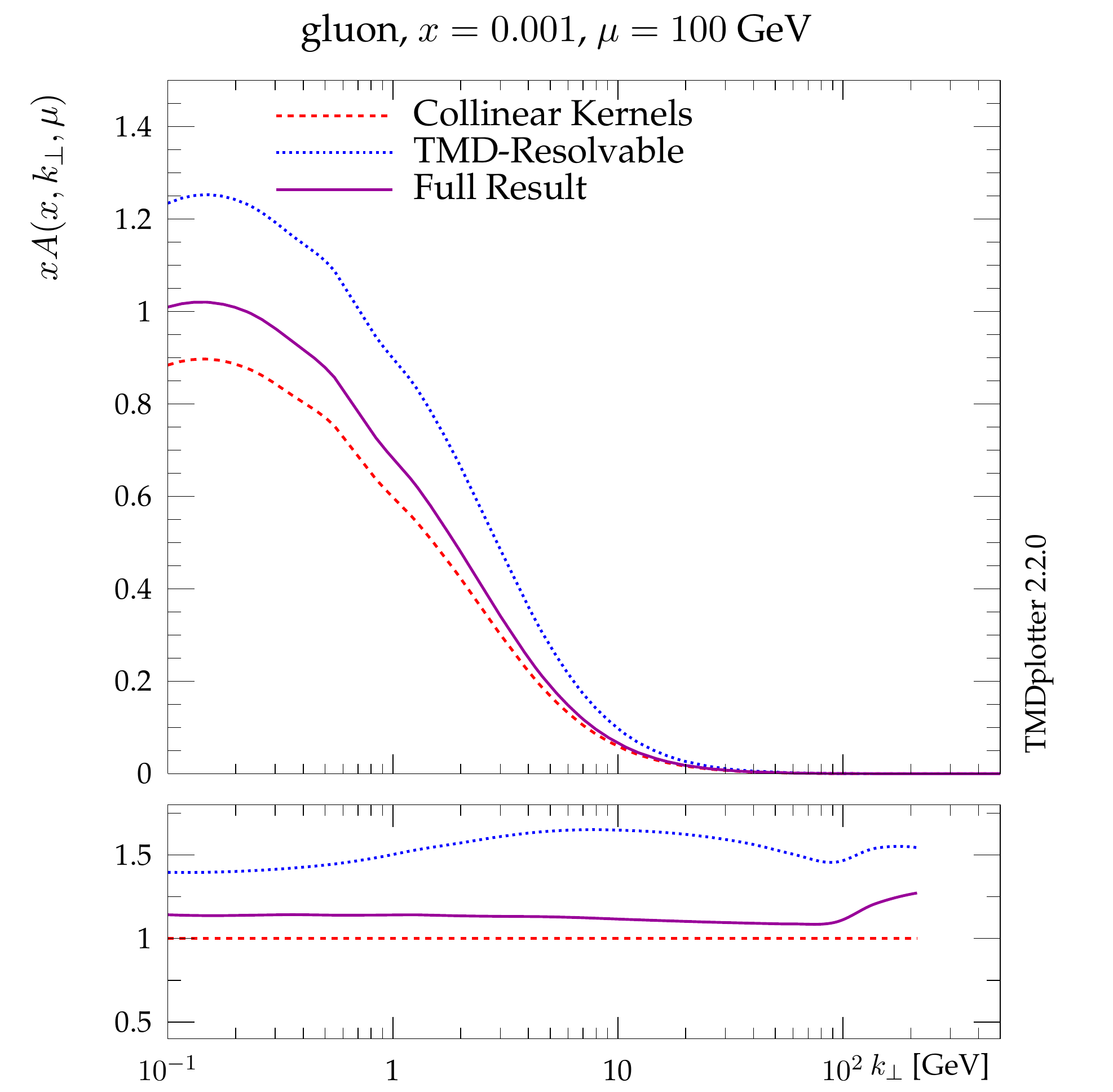}
    }
    \qquad
  \caption{\it  Evolution of the  integrated (top) and  TMD (bottom) 
  parton distributions  
  using the new branching equations with 
  $k_\perp$-dependent splitting kernels. 
  The solid magenta curves give the full result,  
  while the dotted blue curves show the 
  contribution to the full result from  $k_\perp$-dependent splittings in 
  resolvable emissions only.  The dashed red curves show the result 
  of evolution with purely collinear splitting kernels.}
\label{fig2:evolveddistributions}
\end{center}
\end{figure} 

We next present numerical results from 
this MC solution of the branching evolution equations. For the purpose of 
illustrating the implementation of the branching evolution, we can take, 
 as   initial TMD distributions at scale $\mu_0$, 
any of the parameterizations available e.g.~in the 
library~\cite{Abdulov:2021ivr,Hautmann:2014kza}. The numerical 
results which  follow  are obtained by taking  
 the parameterization given 
 in~\cite{Abramowicz:2015mha,BermudezMartinez:2018fsv} 
at  $\mu_0 = 1.4$ GeV.   In 
  Fig.~\ref{fig2:evolveddistributions},    the solid magenta curves
 show  the  results for gluon and down-quark TMD 
distributions evolved to $\mu= 100 $ GeV. In the top panels of 
 Fig.~\ref{fig2:evolveddistributions} we plot    the $x$ dependence of the distributions  integrated over $k_{\perp}^2$, $f_a(x,\mu^2)$, while in the bottom panels  
we plot the $k_\perp$ dependence for fixed $x$. 

  Fig.~\ref{fig2:evolveddistributions} 
presents, besides the solid magenta curves,  two other sets of curves which 
are obtained with the same initial TMD distributions at scale $\mu_0$ but different 
evolution kernels, and 
are displayed  for the purpose of comparison:  the dashed red curves show the 
results which are obtained without including any  $k_\perp$ dependence  in the splitting kernels, that is, with 
the purely collinear splitting kernels; the dotted blue curves show the results 
which are obtained by including 
the $k_\perp$ dependence of splitting functions in resolvable emissions only. 

The comparison of the full result (solid magenta) with the purely collinear result 
(dashed red)  in Fig.~\ref{fig2:evolveddistributions} illustrates that 
   the influence of the TMD splitting 
kernels on evolution is significant especially for  low $x$. 
More  particularly,  it illustrates that the impact of the TMD splittings is not 
washed out by $k_\perp$-integration, and persists 
 at the level of the distributions  integrated over $k_\perp$ as well. 
We stress that the TMD distributions corresponding to 
the solid magenta curves and dashed red curves  both 
fulfill the integral relations  in Eq.~(\ref{eq:ap:conservation}).  
The differences between them 
 stem from dynamical contributions encoded in the TMD splitting functions. 
 These  give rise to 
 a change in the  $k_\perp$  and $x$ shapes of the TMD distributions after evolution.  

The comparison of the full result (solid magenta) with the result from 
including TMD splitting functions in resolvable emissions only (dotted blue) 
 in Fig.~\ref{fig2:evolveddistributions} 
illustrates  the decomposition of TMD splitting effects into  resolvable 
and non-resolvable components.   
As implied by the analysis leading to Eq.~(\ref{eq:tmdPevolktdepSud}), 
the TMD distributions corresponding to 
the  dotted blue curves  in Fig.~\ref{fig2:evolveddistributions}   
do  not fulfill the integral relations in Eq.~(\ref{eq:ap:conservation}).   
Unlike the 
case  described in the previous paragraph, 
therefore, the  departure of the dotted blue curves from the full result  
 can be attributed to violations of  the   relation (\ref{eq:ap:conservation}).

\begin{table}[h]
\centering
\begin{tabular}{l|ccc}
       & \multicolumn{3}{l}{Full Result}                          \\ \hline
$\mu^2$ (GeV$^2$)     & $\alpha_s(\mu^2)$, fix. $z_M$ & $\alpha_s(q_\bot^2)$, fix. $z_M$ & $\alpha_s(q_\bot^2)$, dyn. $z_M$ \\ \hline
%3      & 0.999518       & 0.999493        & 0.999541        \\
%10     & 0.998872       & 0.998736        & 0.998986        \\
%100    & 0.996833       & 0.99591         & 0.997059        \\
%1000   & 0.994076       & 0.991464        & 0.994185        \\
%10000  & 0.990895       & 0.987133        & 0.990727        \\
%100000 & 0.98432        & 0.978276        & 0.983029
3      & 1.000       & 1.000        & 1.000        \\
10     & 0.999       & 0.999        & 0.999        \\
$10^2$    & 0.997       & 0.996        & 0.997        \\
$10^3$   & 0.994       & 0.992        & 0.994        \\
$10^4$  & 0.991       & 0.987        & 0.991        \\
$10^5$ & 0.984       & 0.978        & 0.983        \\ \hline \hline
       & \multicolumn{3}{l}{TMD-Resolvable}                 \\ \hline
$\mu^2$ (GeV$^2$)     & $\alpha_s(\mu^2)$, fix. $z_M$ & $\alpha_s(q_\bot^2)$, fix. $z_M$ & $\alpha_s(q_\bot^2)$, dyn. $z_M$ \\ \hline
%3      & 1.02853        & 1.03814         & 1.00011         \\
%10     & 1.08716        & 1.13884         & 1.00733         \\
%100    & 1.15571        & 1.30351         & 1.04542         \\
%1000   & 1.1949         & 1.41256         & 1.09141         \\
%10000  & 1.21899        & 1.47808         & 1.12848         \\
%100000 & 1.22937        & 1.50695         & 1.14833         \\ \hline \hline
3      & 1.029        & 1.038         & 1.000         \\
10     & 1.087        & 1.139         & 1.007         \\
$10^2$    & 1.156        & 1.304         & 1.045         \\
$10^3$   & 1.195        & 1.413         & 1.091         \\
$10^4$  & 1.219        & 1.478         & 1.129         \\
$10^5$ & 1.229        & 1.507         & 1.148         \\ \hline \hline
       & \multicolumn{3}{l}{Collinear Kernels}                 \\ \hline
$\mu^2$ (GeV$^2$)     & $\alpha_s(\mu^2)$ fix. $z_M$ & $\alpha_s(q_\bot^2)$, fix. $z_M$ & $\alpha_s(q_\bot^2)$, dyn. $z_M$ \\ \hline
%3      & 0.999512       & 0.99956         & 0.999554        \\
%10     & 0.998932       & 0.998866        & 0.999025        \\
%100    & 0.997148       & 0.99656         & 0.997291        \\
%1000   & 0.994756       & 0.993228        & 0.99476         \\
%10000  & 0.992024       & 0.989441        & 0.991737        \\
%100000 & 0.985687       & 0.981129        & 0.983851        \\ \hline \hline
3      & 1.000       & 1.000         & 1.000        \\
10     & 0.999       & 0.999         & 0.999        \\
$10^2$    & 0.997       & 0.997         & 0.997        \\
$10^3$   & 0.995       & 0.993         & 0.995         \\
$10^4$  & 0.992       & 0.989         & 0.992        \\
$10^5$ & 0.986       & 0.981         & 0.984       
\end{tabular}
\caption{\it Momentum sum rule check 
for (top) full result; (middle)  result from  $k_\perp$-dependent splittings 
in   resolvable emissions only; (bottom)  result 
  from purely collinear splitting kernels. The three columns correspond 
to three different boundary conditions 
on the strong coupling  $\alpha_s$ and 
  soft-gluon resolution scale $z_M$.}
\label{tab1:sumruletable}
\end{table}

To examine the size of these violations,
in Tab.~\ref{tab1:sumruletable} we perform a numerical check of
 momentum sum rules.
We present numerical results for $\sum_a\int_{x_0}^1 dx \ x f_a(x,\mu^2)$, where $x_0$ is a small fixed value. 
We  have verified numerically the convergence of the result 
 for decreasing $x_0$ and for the results in the Table we use $x_0 = 10^{-5}$. 
We report results for 
  the three cases  considered   in Fig.~\ref{fig2:evolveddistributions}, at different
 values of $\mu$.  The three columns for each
 case  correspond  to   different scenarios  for the scale of the running coupling $\alpha_s$ and
 the soft-gluon resolution scale $z_M$:
 the first two have constant $z_M$ (``fix. $z_M$")~\cite{BermudezMartinez:2018fsv}, one 
with $\alpha_s$ taken at the branching scale $\mu$
 and one with $\alpha_s$ taken at the transverse momentum $q_\perp$, while the third has
 $\mu$-dependent (dynamical)  $z_M$ (``dyn. $z_M$")~\cite{Hautmann:2019biw}.  Consistently with the 
 discussion above, 
 the table shows that, regardless of the 
 scenario for $\alpha_s$ and $z_M$, the  momentum sum rule is satisfied, within the numerical accuracy, 
in the case of  the branching evolution with TMD splitting functions  proposed in this 
paper (top rows) as well as in the case of  purely collinear splitting functions (bottom rows), while it 
is violated  in the case of TMD splitting functions in real-emission kernels and collinear Sudakov factors 
(middle rows). 

The numerical implementation of an approach 
which both includes the TMD splitting functions and 
satisfies the momentum sum rule is one of the main 
achievements of this paper. 
The construction of a full MC event generator 
implementing this  approach,  
e.g. using the methods of~\cite{Baranov:2021uol},   
will be the subject of future work. 
Such a MC could be compared with 
existing small-$x$ MC generators based on 
high-energy factorization~\cite{Catani:1990eg}, 
e.g.~\cite{Jung:2010si,Andersen:2011hs,Chachamis:2015zzp,vanHameren:2016kkz}.    

In conclusion,  we have formulated a parton branching 
which is 
applicable to TMD and integrated  distributions and 
for the first time takes into account 
both the $z$ and $k_\perp$ dependence 
of splitting functions defined from the high-energy limit 
of partonic decay processes. These (off-shell) splitting functions 
have well-prescribed collinear and high-energy limits: they 
coincide with the customary leading-order collinear splitting functions 
for $k_\perp \to 0$,  and contain finite-$k_\perp$ corrections 
which are responsible for the all-order resummation of logarithmically enhanced 
contributions to parton evolution for  $x \to 0$.  
We have introduced new Sudakov form factors, for both gluon and quark 
channels,  depending on the angular-averaged TMD splitting functions. 
Using these elements, our approach describes  
  resolvable and non-resolvable branchings. 
We have presented its  MC implementation, 
 and applied it to obtain numerical results  
for  the evolution of TMD distributions, 
and to verify explicitly integral relations expressing 
the momentum sum rules.

\vskip 0.9 cm 

\noindent {\bf Acknowledgments}. We thank H.~Jung for useful discussions.  
M.~H. gratefully acknowledges support by Consejo Nacional de Ciencia 
y Tecnolog\'ia grant number A1 S-43940 (CONACYT-SEP Ciencias B\'asicas).
A.~K. is grateful for the support of Narodowe Centrum  
Nauki under grant SONATAbis no 2019/34/E/ST2/00186.   
 K.~K.  acknowledges the European 
Union's Horizon 2020 research and innovation programme  
under grant agreement No. 824093. A.~L. acknowledges funding 
by Research Foundation-Flanders (FWO) (application number: 1272421N). 
This work is supported partially by grant V03319N from the common 
FWO-PAS exchange program.\\

%\appendix 

\end{document}